# Review of anisotropic terahertz material response

Takashi Arikawa*, Qi. Zhang, Lei Ren, Alexey A. Belyanin, Junichiro Kono

Abstract

Anisotropy is ubiquitous in solids and enhanced in low-dimensional materials.  In response to an electromagnetic wave, anisotropic absorptive and refractive properties result in dichroic and birefringent optical phenomena both in the linear and nonlinear optics regimes.  Such material properties have led to a diverse array of useful polarization components in the visible and near-infrared, but mature technology is non-existent in the terahertz (THz).  Here, we review several novel types of anisotropic material responses observed in the THz frequency range, including both linear and circular anisotropy, which have long-term implications for the development of THz polarization optics.  We start with the extreme linear anisotropy of macroscopically aligned carbon nanotubes, arising from their intrinsically anisotropic dynamic conductivity.  Magnetically induced anisotropy will then be reviewed, including the giant Faraday effects observed in semiconductors, semimetals, and two-dimensional electron systems.

*T. Arikawa · Q. Zhang · L. Ren · J. Kono*
*Department of Electrical & Computer Engineering, Rice University,*
*Houston, TX 77005, USA*
*e-mail: kono@rice.edu*

*A. A. Belyanin*
*Department of Physics, Texas A&M University,*
*College Station, TX 77843, USA*

*\*Present address Department of Physics, Kyoto University,*
*Kyoto, 606-8502, Japan*

# 1 Introduction

Anisotropic optical response of materials has been extensively studied and exploited for years since the discovery of birefringence in uniaxial calcite crystals [1]. This intrinsic anisotropy led to deeper understanding of material properties as well as various applications in diverse fields of research and industry. One of the most important applications is polarization optics, such as wave plates and polarizing beam splitters with high extinction ratios, which enable us to perform very precise and advanced optical measurements.

In contrast to the technical maturity in the visible region, much room remains for the research and application of anisotropic material response in the terahertz (THz) frequency range. Grischkowsky *et al*. measured refractive indexes of sapphire and quartz for ordinary and extraordinary rays in the THz, showing strong birefringence [2]. Masson *et al*. designed and demonstrated a THz achromatic (0.25–1.75 THz) quarter-wave plate using six quartz plates [3]. Liquid crystals also exhibit strong birefringence due to their orientation order, which can be controlled by an external electric or magnetic field [4-6, Park, Hsieh]. Artificial materials such as metamaterials with low symmetry also show anisotropic response and have been used to demonstrate THz polarization control [7]. On the other hand, by taking advantage of the long wavelength of THz waves compared to visible light, Scherger *et al*. demonstrated a very simple wave plate made of ordinary office paper [8]. Interestingly, even woods possess some birefringence in the THz frequency range due to preferential fiber orientation [9].

In addition to intrinsic anisotropy of crystal structures, anisotropy can be *induced* in an extrinsic manner by the application of an external field, e.g., an electric or magnetic field, leading to an electro-optical (EO) or magneto-optical (MO) effect, respectively. These effects are also diversely used in optoelectronic applications, including EO sampling of coherent THz pulses. In EO effects (e.g., the Pockels and Kerr effects), the induced birefringence is similar to intrinsic birefringence in anisotropic crystals, i.e., the system behaves differently in response to ordinary and extraordinary rays. On the other hand, MO effects (e.g., the Faraday and MO Kerr effect) usually induce a difference in refractive index between left- and right-circularly polarized light; this circular birefringence and dichroism come from the finite off-diagonal elements of the dielectric tensor.

In the THz frequency region, the off-diagonal components often stem from cyclotron motion of free carriers under magnetic field [10-13].

In this article, we review recent results, of our own as well as other groups, on intrinsic and extrinsic anisotropy observed in the response of materials to THz waves.  First, we will briefly summarize how the anisotropic response can be macroscopically described by a dielectric (or equivalently, conductivity) tensor. Next, as a prime example of intrinsic anisotropy, we describe results of polarization-dependent THz transmission measurements on films of macroscopically-aligned carbon nanotubes (CNTs), whose orientation order is far more perfect than that of fibers in wood and paper mentioned above [14-17, 18-19].  The results show their excellent performance as linear polarizers.  Next, we present free-carrier Faraday effects observed in several material systems as examples of extrinsic anisotropy in the THz.  We describe giant Faraday effects observed in $n$-InSb (a narrow-gap semiconductor), HgTe (a semimetal), and graphene (a zero-gap semiconductor), which are promising material systems for broadband THz polarization optics and modulators [20-22].  Finally, studies of polarization-dependent THz magneto-optical spectroscopy of two-dimensional electron gases (2DEG) in modulation-doped GaAs quantum wells are presented, including the observation of a quantum Hall plateau in AC Hall conductivity [23] and the demonstration of coherent control of cyclotron resonance (CR) [24].

## 2 Phenomenological description of anisotropic material response

In this section, we will briefly give a phenomenological description of linear and circular anisotropy.  Macroscopically, the response of a material to an electromagnetic wave is described by its dielectric tensor, $\varepsilon_{ij}$ ($i,j = x, y, z$).  In three-dimensional crystalline solids, the number of non-vanishing and independent components of the dielectric tensor is determined by the crystal structures.  For uniaxial crystal structures, for example, the dielectric tensor is diagonal in the principal dielectric coordinate system and only the $\varepsilon_{zz}$ component is different from the other two,

$$\varepsilon = \begin{pmatrix} \varepsilon_{xx} & 0 & 0 \\ 0 & \varepsilon_{xx} & 0 \\ 0 & 0 & \varepsilon_{zz} \end{pmatrix}. \qquad (1)$$

From the Maxwell equations, the normal modes that the uniaxial medium can support are two orthogonal linear polarizations with different propagation constants, which lead to linear birefringence and dichroism [25]. For more details see, e.g., [26]. In the case of low-dimensional materials, the electromagnetic anisotropy is significantly enhanced due to the intrinsic structural anisotropy of the medium (e.g. between in-plane and out-of-plane direction in a thin film) or anisotropy of constituent nanoparticles. For example, one-dimensional CNTs absorb light when the light polarization is parallel to the CNT axis, but when the light polarization is perpendicular to the CNT axis, there is almost no interaction (Section 3). If one has a collection of CNTs that are macroscopically aligned, its effective dielectric tensor retains the individual anisotropy and shows extreme linear dichroism.

To describe the Faraday effect, let us consider the effective dielectric tensor of an isotropic medium in the Faraday geometry, i.e., both the external magnetic field and the propagation direction of the electromagnetic wave are along the $z$ axis. From the symmetry consideration, i.e., $\varepsilon_{ij}$ is invariant under any rotation of the coordinate axis about the $z$ axis, the effective dielectric tensor takes the gyrotropic form,

$$\varepsilon = \begin{pmatrix} \varepsilon_{xx} & \varepsilon_{xy} & 0 \\ -\varepsilon_{xy} & \varepsilon_{xx} & 0 \\ 0 & 0 & \varepsilon_{zz} \end{pmatrix}. \qquad (2)$$

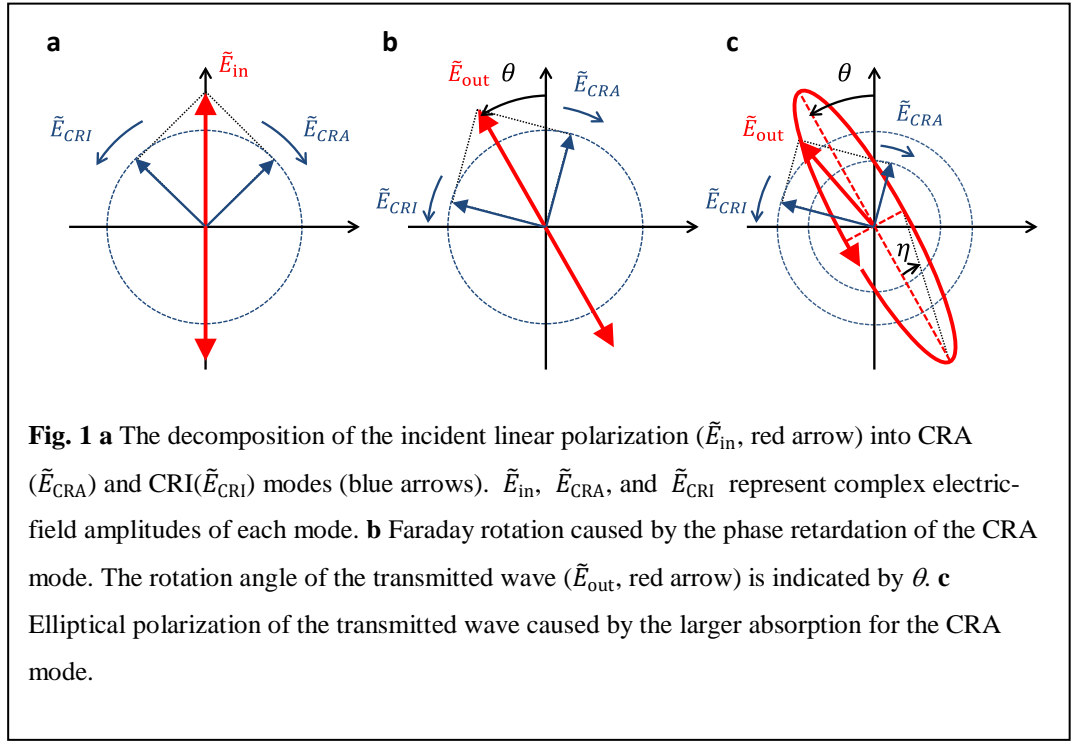

**Fig. 1 a** The decomposition of the incident linear polarization ($\tilde{E}_{in}$, red arrow) into CRA ($\tilde{E}_{CRA}$) and CRI($\tilde{E}_{CRI}$) modes (blue arrows). $\tilde{E}_{in}$, $\tilde{E}_{CRA}$, and $\tilde{E}_{CRI}$ represent complex electric-field amplitudes of each mode. **b** Faraday rotation caused by the phase retardation of the CRA mode. The rotation angle of the transmitted wave ($\tilde{E}_{out}$, red arrow) is indicated by $\theta$. **c** Elliptical polarization of the transmitted wave caused by the larger absorption for the CRA mode.

From the Maxwell equations, the normal modes in this medium are two oppositely rotating circular polarizations with different refractive indexes, which lead to circular anisotropy and dichroism [10]. In the free-carrier Faraday effect, the normal modes are called cyclotron resonance active (CRA) or inactive (CRI) modes, depending on whether or not the mode can couple to the cyclotron motion of the free carriers. The propagation constants are given as follows,

$$\tilde{n}^2_{CRA(CRI)} = \varepsilon_{xx} \pm i\varepsilon_{xy}. \qquad (3)$$

where the CRA/CRI notation corresponds to the coupling to electrons. The components of the dielectric tensor are modeled by various microscopic theories.

Figure 1 shows how the Faraday effect can be understood from the different complex refractive indexes for CRA and CRI modes. We start by decomposing the incident electromagnetic wave with linear polarization into CRA and CRI modes with the same amplitude and phase (Fig. 1a). The difference in the refractive index between the CRA and CRI modes (here we assume Re[$\tilde{n}_{CRA}$] > Re[$\tilde{n}_{CRI}$]) produces phase retardation of the CRA mode, which results in the polarization rotation of the transmitted wave (Fig. 1b). The additional difference in the absorption coefficient results in the amplitude difference and makes the transmitted wave elliptically polarized (Fig 1c). The following Faraday parameters (rotation angle $\theta$ and ellipticity $\eta$) characterize the Faraday effect [10],

$$\theta = \frac{\arg(\tilde{E}_{CRA}) - \arg(\tilde{E}_{CRI})}{2}, \quad (4)$$

$$\eta = \frac{|\tilde{E}_{CRI}| - |\tilde{E}_{CRA}|}{|\tilde{E}_{CRI}| + |\tilde{E}_{CRA}|}. \quad (5)$$

An alternative definition of the ellipticity in the radian unit is shown in the Fig 1c.

# 3 Anisotropic THz response of macroscopically aligned carbon nanotubes

Carbon nanotubes (CNTs), with extremely high length-to-diameter ratios of up to $10^8$, show unprecedentedly strong anisotropy in electric, magnetic, and optical properties [18, 27]. Individual metallic single-wall carbon nanotubes (SWCNTs) have been shown to be excellent 1-D electrical conductors [28], while individual semiconducting SWCNTs have been shown to absorb light only when the light polarization is parallel to the tube axis [29]. To take full advantage of their excellent 1-D properties as polarization devices, several methods have been developed for achieving a high degree of alignment of SWCNTs on macroscopic scales. Here we show extremely anisotropic response of films of macroscopically aligned ultralong CNTs in the THz frequency range and discuss their quality as THz linear polarizers [14-19].

## 3.1 Highly aligned single-wall carbon nanotubes

The aligned SWCNT films were obtained by vertical growth from 1 to 2 μm wide pads of catalyst (0.5 nm of Fe and 10 nm of $Al_2O_3$) defined by optical lithography and deposited using electron-beam evaporation (Fig. 2a), followed by SWCNT growth in a hot-filament furnace by water-assisted CVD [30–33], as detailed in [34–36]. The length of the nanotubes was controlled by the exposure time, and their average diameter was around 2.7 nm. The highly aligned nanotubes were then transferred onto sapphire substrates (Fig. 2b), producing homogeneous, $cm^2$-large films of highly aligned SWCNTs, with a thickness of ≈ 2 μm. Figure 2c shows a scanning electron microscope image showing the excellent SWCNT alignment on the transferred substrate.

Figure 2d shows polarization-dependent transmittance spectra for a typical aligned SWCNT film. The transmission is very small when the THz polarization is parallel to the SWCNT alignment direction, whereas there is nearly complete transmission in the perpendicular case. The THz transmittance in the parallel configuration decreases with increasing frequency, unlike the commercial wire-grid THz polarizers. This peculiar frequency dependence, coming from a collective antenna effect [17], is a promising feature to realize THz polarizers

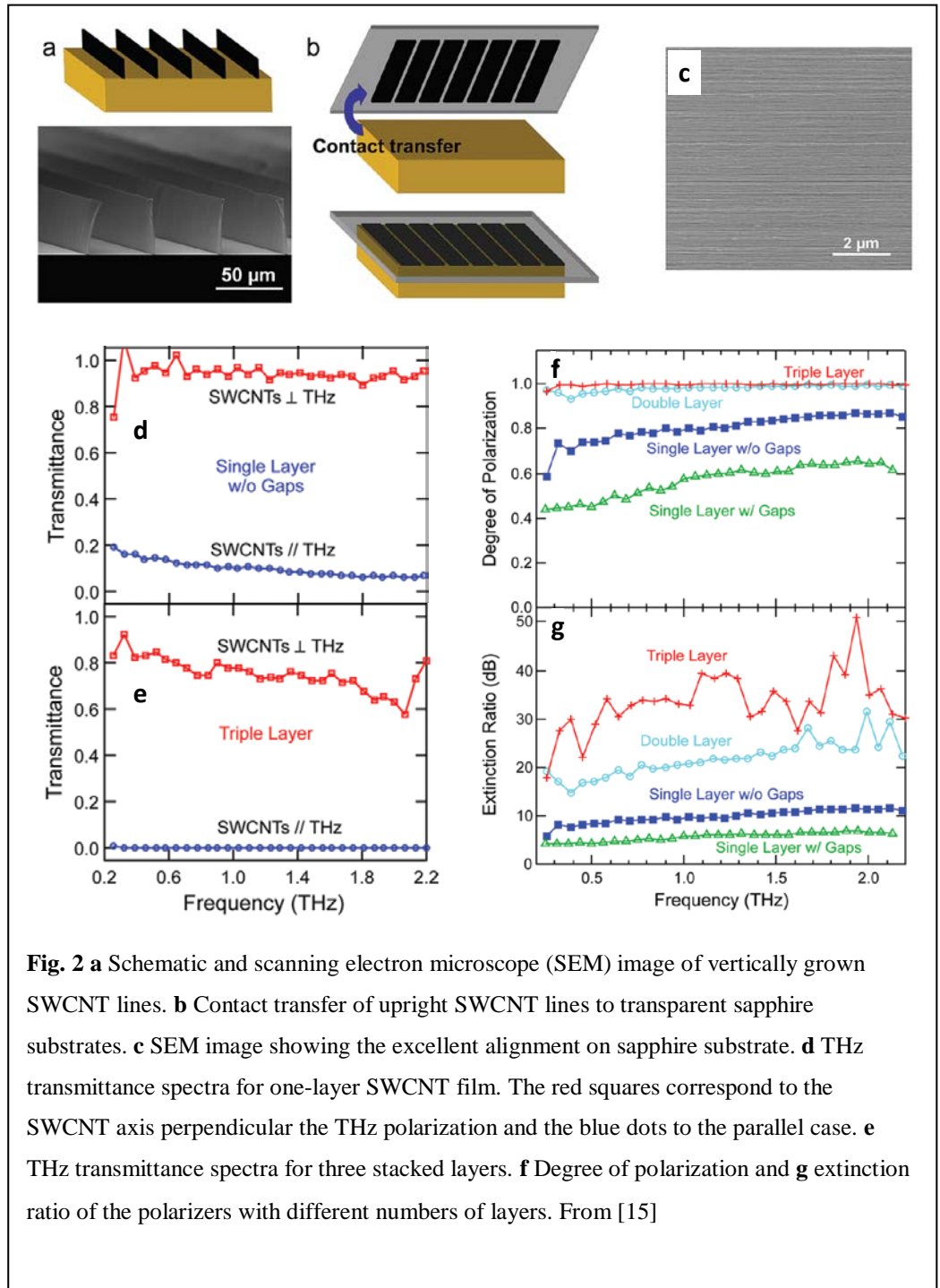

**Fig. 2 a** Schematic and scanning electron microscope (SEM) image of vertically grown SWCNT lines. **b** Contact transfer of upright SWCNT lines to transparent sapphire substrates. **c** SEM image showing the excellent alignment on sapphire substrate. **d** THz transmittance spectra for one-layer SWCNT film. The red squares correspond to the SWCNT axis perpendicular the THz polarization and the blue dots to the parallel case. **e** THz transmittance spectra for three stacked layers. **f** Degree of polarization and **g** extinction ratio of the polarizers with different numbers of layers. From [15] working in the higher frequency region, which is hard to achieve with wire-grid technology. When three layers of ~2-μm thick aligned SWCNT films are stacked, the transmission drops down to almost zero in the whole frequency range for the parallel configuration but remains very close to one in the perpendicular case (Fig. 2e).

For the quantitative characterization of these devices as linear polarizers, the degree of polarization (DOP) and extinction ratio (ER), respectively defined as DOP = $(T_\perp - T_{//})/(T_\perp + T_{//})$ and ER = $T_{//}/T_\perp$ were calculated (Fig. 2f,g). Both the DOP and ER strongly increases with the number of layers in the whole frequency range, reaching values of 0.999 and 33.4 dB, respectively, for triple layers. These results clearly demonstrate that highly aligned SWCNT films perform as ideal linear polarizers in the THz frequency range.   The insertion loss is mainly due to the reflection from the substrate, which can be addressed by, e.g., an anti-reflection coating on the substrate.   The working frequency range could be extended up to the mid-infrared [37] and visible [36] to design ultrawide spectral absorbers.

## 3.2 Reel-wound multi-wall carbon nanotubes

Kyuoung *et al*. demonstrated a simple method to fabricate CNT linear polarizers [19].   They drew CNT sheets from a multi-wall carbon nanotube forest and wound on a U-shaped reel, which produced aligned, freestanding micrometer-thick linear polarizers (Fig. 3a,b).   A typical time-domain THz spectroscopy system was used to characterize the performance of the CNT linear polarizer, and that of a commercial wire-grid polarizer was compared.   Figure 3c (3d) shows ER (DOP) of the CNT and commercial wire-grid linear polarizer, which shows better performance of the CNT linear polarizer.    Especially in the higher frequency region, ER (DOP) of around 37 dB (0.999) is achieved for the CNT linear polarizer, whereas the performance of the wire-grid linear polarizer is inevitably getting worse.   The only issue seems to be an insertion loss of several tens of percent, presumably due to the imperfect CNT alignment, but the easy fabrication of reel-wound CNT polarizers is attractive and promising for commercialization.

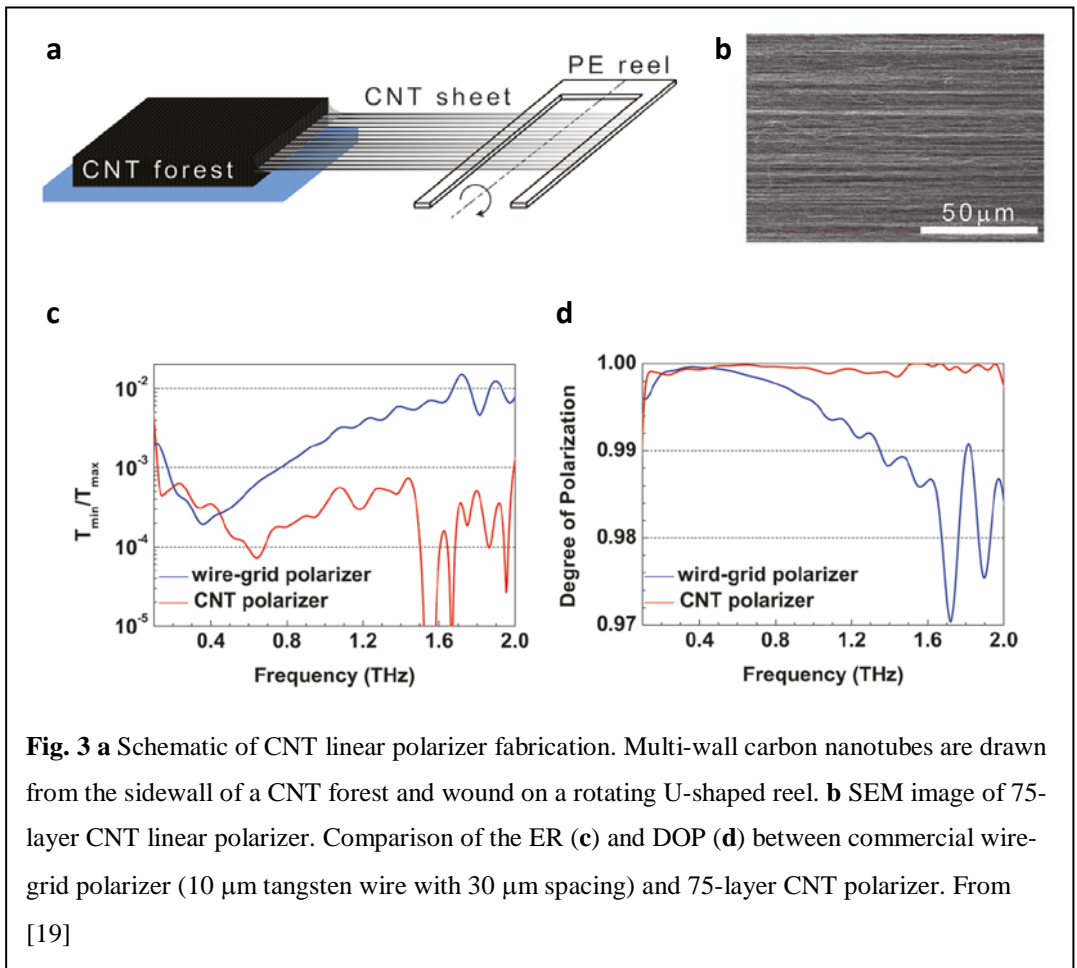

**Fig. 3 a** Schematic of CNT linear polarizer fabrication. Multi-wall carbon nanotubes are drawn from the sidewall of a CNT forest and wound on a rotating U-shaped reel. **b** SEM image of 75-layer CNT linear polarizer. Comparison of the ER (**c**) and DOP (**d**) between commercial wire-grid polarizer (10 μm tangsten wire with 30 μm spacing) and 75-layer CNT polarizer. From [19]

# 4 Giant Faraday effect in narrow-gap semiconductors and semimetals

The application of a magnetic field breaks the time-reversal symmetry, which introduces finite off-diagonal elements in the dielectric tensor and makes the system circularly dichroic. In the case of free carriers, the cyclotron motion of electrons or holes selectively couples to circularly polarized light with opposite senses, CRA and CRI mode [38]. The anisotropy induced between the complex indices of refraction for CRA and CRI modes leads to strong Faraday effects. Narrow-gap semiconductors and semimetals show giant free-carrier Faraday effect in the THz frequency range [20-22]. Here, we review recent reports on the giant Faraday effects observed in *n*-InSb, HgTe, and grapheme in the THz frequency range and discuss their potential as THz polarization optics, such as circular polarizers, wave plates, isolators, and modulators.

## 4.1 *n*-InSb

InSb, which is a narrow-gap semiconductor, has been known to show a giant free-carrier Faraday effect in the microwave and far-infrared range since the 1960s [39]. Here we describe our recent results on the broadband THz Faraday effect in *n*-InSb by THz time-domain spectroscopy and discuss its potential as tunable and broadband polarization optics in the THz frequency region [20].

  A Te-doped <111> *n*-InSb crystal (20 mm × 30 mm × 0.63 mm) with a doping electron density of $6.1 \times 10^{14}$ cm$^{-3}$ was used. The sample temperature was kept at 184 K so that the THz response was dominated by free carriers rather than bound carriers. The electrons were thermally excited and the total electron density was ~$7.6 \times 10^{14}$ cm$^{-3}$ [40]. Coherent THz pulses with elliptical polarization were generated from a two-color laser induced plasma [41]. The first wire-grid polarizer was placed before focusing the THz wave to the sample to make the incident THz wave linearly polarized along the *x* axis. The transmitted THz electric field through the sample was measured by an electro-optic sampling method using a (110) ZnTe crystal (1 mm thick) [42]. Two different detection geometries (angles of the second wire-grid polarizer (analyzer) after the sample and ZnTe crystal) were used to separately measure $x$- and $y$-components of the transmitted THz pulses, as detailed in [20].

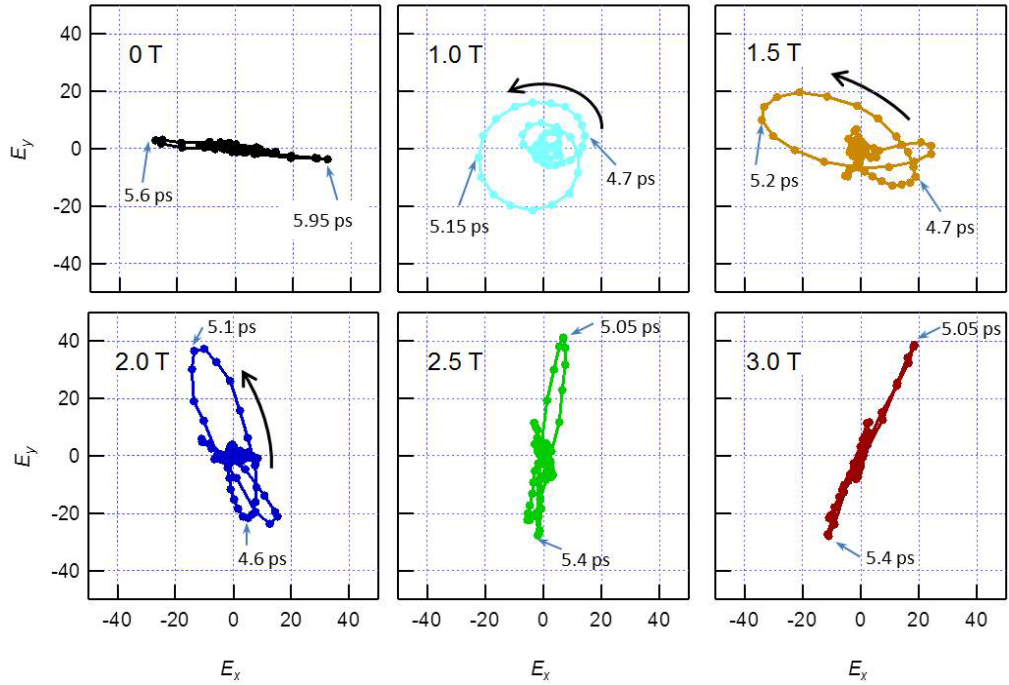

**Fig. 4** Parametric plots of *x*-, and *y*-components of the transmitted THz pulses. Each point represents the tip of the electric field vector at each time. The time interval between the dots is 50 fs. The polarization of the transmitted THz pulses changes drastically depending on the external magnetic field, demonstrating a giant Faraday effect. From [20]

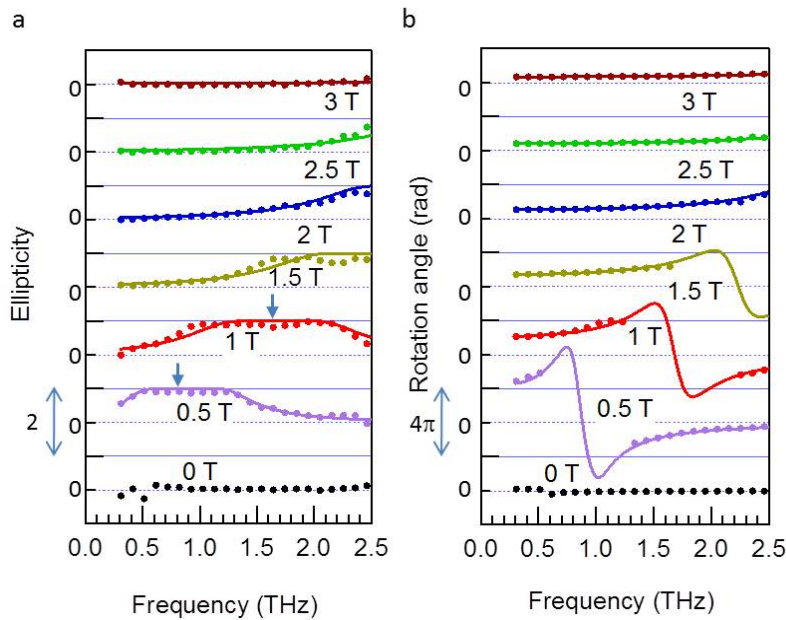

**Fig. 5** Faraday parameters. The dots show experimental data and the solid lines show fitting results with the cold magneto-plasma model. a) The magnetic field dependence of the ellipticity. The down arrows show the CR frequency at each magnetic field. b) The magnetic field dependence of the rotation angle. In the frequency region where the ellipticity is unity, the rotation angle cannot be determined. From [20].

A drastic change was observed in the polarization state after transmission through the sample in a magnetic field. The parametric plots of the transmitted THz pulses (Fig. 4) show that the polarization of the incident pulse changes from $x$-polarization to almost circular polarization (~1 T), elliptical polarization (1.5 ~ 2.0 T), and tilted linear polarization (> 2.5 T). The Faraday parameters were deduced from the data in Fig. 4 (Fig. 5). Below 2.5 T, Faraday ellipticities reach 1 around the CR frequencies $\omega_c = 2\pi f_c = eB/m_e$, where $e$ is the electronic charge, $m_e \approx 0.018 m_0$ is the effective mass of electrons, and $m_0$ is the free electron mass. This means that the transmitted light is purely the CRI mode because the CRA mode is completely absorbed by free carriers (see Equation (5)). This demonstrates that $n$-InSb works as a broadband (from 0.9 to 2.3 THz at 1 T) circular polarizer with a tunable center frequency with the external magnetic field. At 3 T, the CR frequency is much higher than the experimental frequency window, resulting in the almost zero ellipticity and dispersionless rotation angle. This shows that $n$-InSb also works as a broadband (from 0.3 to 2.5 THz at 3 T) half-wave plate. The rotation angle can be set to $\pi/2$ by changing the thickness of the $n$-InSb crystal and fine-tuning the magnetic field.

The Faraday parameters are excellently reproduced by modeling the dielectric tensor of $n$-InSb with a cold magneto-plasma model (solid lines in Fig. 5) [10, 43-44]. One can determine the electron effective mass and scattering time as fitting parameters, and calculate the complex refractive indexes of $n$-InSb for CRA and CRI modes as shown in Fig. 6a for $B = 1$ T. One can see some dispersion due to CR at ~1.65 THz in the index for the CRA mode, while there is no dispersion in the CRI mode. This clearly shows circular anisotropy induced by the magnetic field.

For device applications, it is very important that the device works without cryogens. The cold magneto-plasma model suggest that at elevated temperatures one can achieve similar performance to that in cooled samples by increasing the magnetic field and decreasing the sample thickness. Figure 6b is plotted for $T = 250$ K (which is achievable with thermoelectric cooler) and the magnetic field $B = 5$ T and a sample of 0.3 mm thickness. Recent advancements in the table-top repetitive pulsed electromagnets with peak magnetic fields of around 10 T [45] will enable us to realize such devices including fast THz polarization modulators.

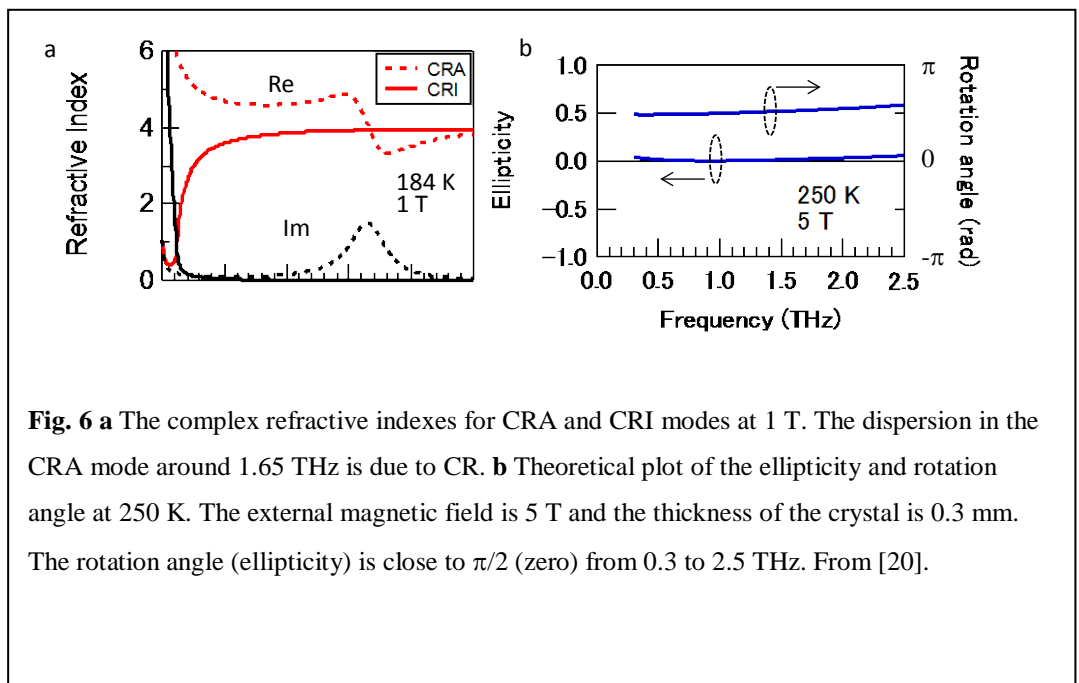

**Fig. 6 a** The complex refractive indexes for CRA and CRI modes at 1 T. The dispersion in the CRA mode around 1.65 THz is due to CR. **b** Theoretical plot of the ellipticity and rotation angle at 250 K. The external magnetic field is 5 T and the thickness of the crystal is 0.3 mm. The rotation angle (ellipticity) is close to $\pi/2$ (zero) from 0.3 to 2.5 THz. From [20].

## 4.2 HgTe

HgTe is a semimetal with a band overlap of 320 meV [46]. A huge number of electrons and holes are thermally excited at room temperature (Fig 7a). In addition, extremely pure, thin film of HgTe available today provides a very high mobility of electrons. These properties are promising for realizing THz polarization optics based on the Faraday effect in HgTe. Shuvaev *et al*. used 70-nm-thick and 1-μm-thick HgTe layers (nominally undoped) grown by molecular beam epitaxy on an insulating CdTe substrate and observed a giant Faraday effect [21]. Figure 7b,c show the magnetic field dependence of Faraday rotation angle angle and ellipticity of HgTe thin films measured at 0.35 THz. The rotation angle of about π/12 rad is obtained for the sample as thin as 70 nm, which corresponds to a large Verdet constant of $3 \times 10^6$ rad $T^{-1}m^{-1}$ as compared to that of InSb (~$10^4$ rad $T^{-1}m^{-1}$). For the 1-μm-thick film, the rotation angle is around π/4 and the ellipticity is small at 3 T, which indicates that this HgTe layer could be used as a Faraday isolator. To identify the origin of this giant Faraday effect, they determined the effective mass, carrier concentration, and mobility as a function of temperature (Fig. 7d-f) by fitting the data using the Drude model. The carrier effective mass varies only slightly with temperature and $m_e = 0.030 m_0$ at 100 K. The value corresponds to the electron effective mass in the light-hole band, which has conduction-band-like character due to the band inversion. This shows that although both electrons and holes are thermally excited, the dominant contribution comes from electrons due to their much smaller effective mass than that of holes in the heavy-hole band. The electron concentration reaches as high as ~ $10^{17}$ $cm^{-3}$ at room temperatures, which is two orders magnitude higher than that of InSb. The mobility decreases at higher temperatures, but it remains high (~$10^4$ $cm^2V^{-1}s^{-1}$) even at room temperatures. These special properties of semimetalic HgTe thin film lead to a giant Faraday effect, which has strong potential for practical applications in THz polarization optics.

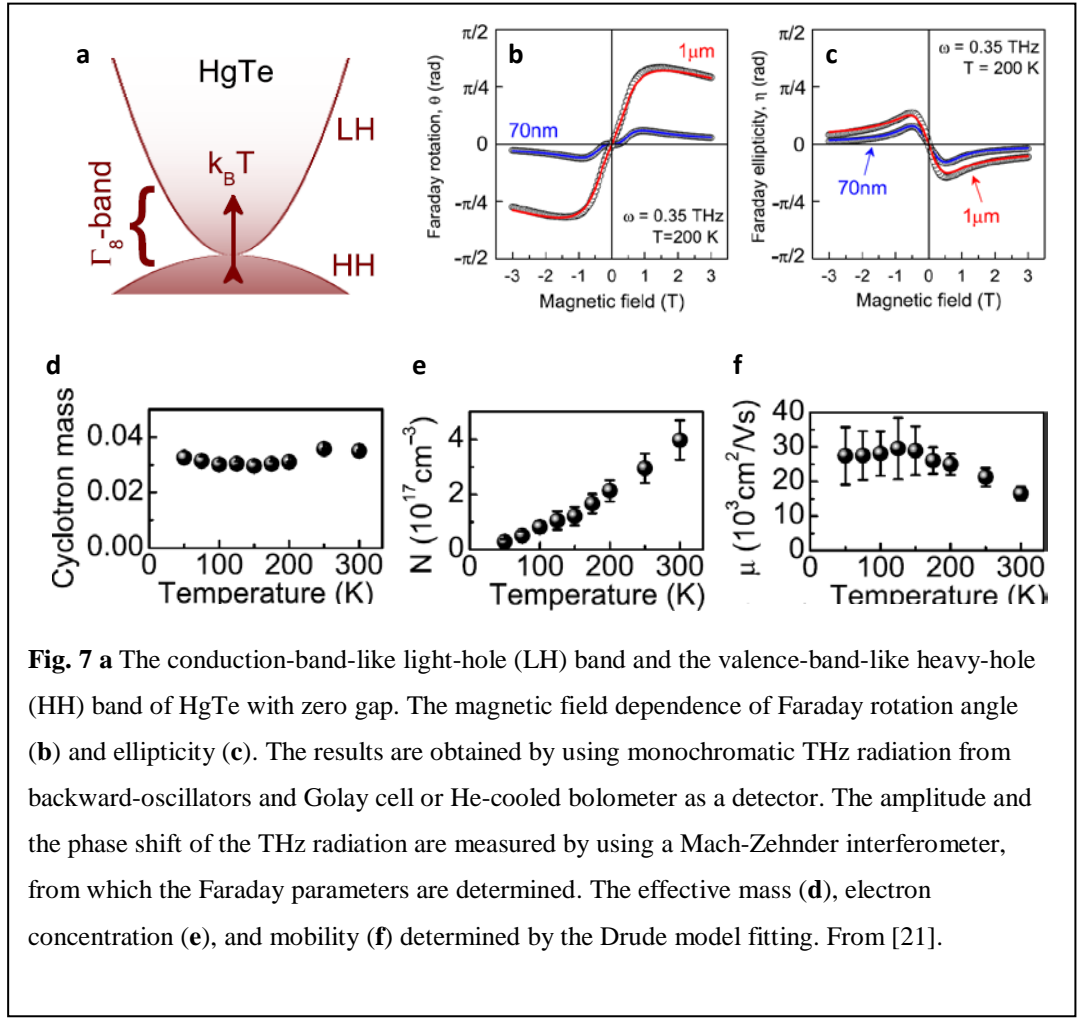

**Fig. 7 a** The conduction-band-like light-hole (LH) band and the valence-band-like heavy-hole (HH) band of HgTe with zero gap. The magnetic field dependence of Faraday rotation angle (**b**) and ellipticity (**c**). The results are obtained by using monochromatic THz radiation from backward-oscillators and Golay cell or He-cooled bolometer as a detector. The amplitude and the phase shift of the THz radiation are measured by using a Mach-Zehnder interferometer, from which the Faraday parameters are determined. The effective mass (**d**), electron concentration (**e**), and mobility (**f**) determined by the Drude model fitting. From [21].

### 4.3 Graphene

Graphene, a single layer of $sp^2$-bonded carbon atoms, is a zero-gap semiconductor with the conduction and valence bands having linear dispersion relations and touching each other at the so-called Dirac point. Crassee *et al*. observed polarization rotation by several degrees in *single layer* graphene, which is colossal given that it comes from an atomically thin material [22]. Figure 8a shows the rotation angle in the THz frequency region at several magnetic fields. The spectra show strong field dependence. In the inset, the rotation angle at 10 and 27 meV are plotted as a function of magnetic field. The data essentially shows a linear trend, which leads to an extremely large Verdet constant as high as $10^7$ rad $T^{-1}$ $m^{-1}$ with the assumption that the graphene thickness is equal to the interlayer distance of graphite. The zero-field-normalized transmission spectra (Fig. 8b) also show strong magnetic field dependence. The inset shows the corresponding absorption spectra at 0 and 7 T, which clearly demonstrate a Drude-like response (which is CR in a magnetic field), respectively.  The results are well fitted with a classical Drude model (dashed lines in Fig. 8).  The Drude weight gives the absolute value of the Fermi energy of 0.35 eV, which agrees well with the estimate from the mid-infrared absorption onset and angle-resolved photoemission spectroscopy.  Most strikingly, the sign of the slope of the rotation spectrum at the CR frequency can unambiguously tell the carrier type.  In the case of the data in Fig. 8a, it is hole CR.  They also observed electron-type Faraday rotation in multilayer graphene, which is highly electron-doped.  The electrical tunability of the Fermi energy in graphene, both in the electron and hole regime, would enable fast polarization modulations in the THz range.

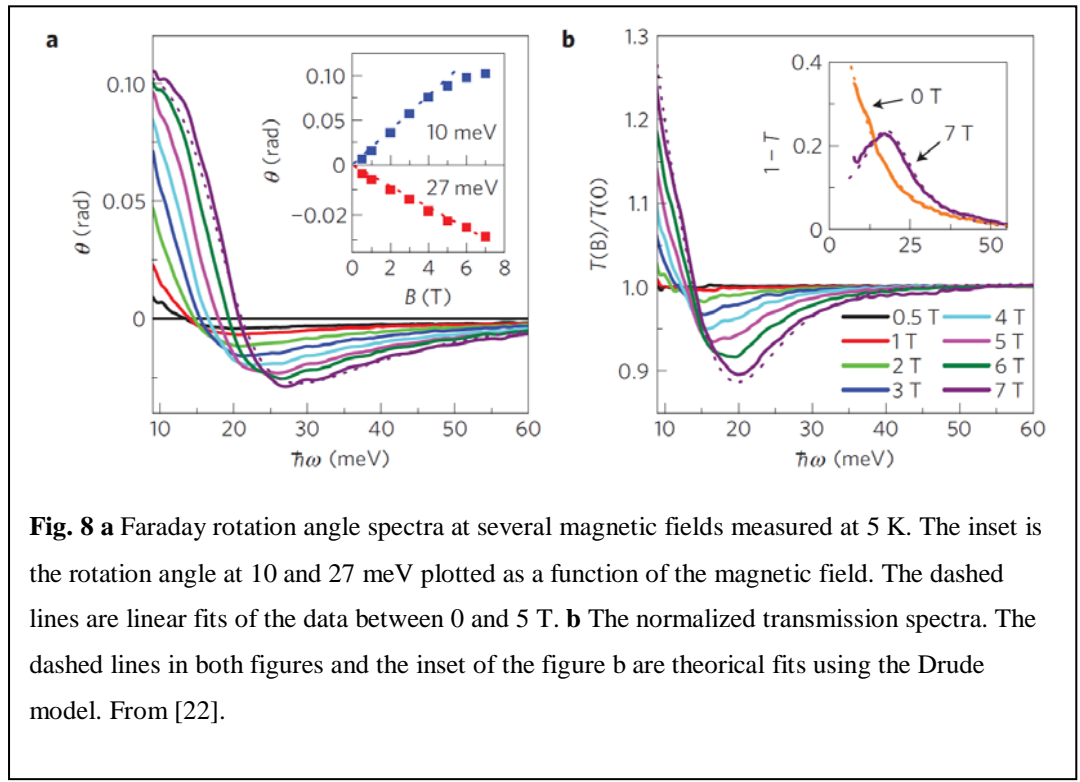

**Fig. 8 a** Faraday rotation angle spectra at several magnetic fields measured at 5 K. The inset is the rotation angle at 10 and 27 meV plotted as a function of the magnetic field. The dashed lines are linear fits of the data between 0 and 5 T. **b** The normalized transmission spectra. The dashed lines in both figures and the inset of the figure b are theorical fits using the Drude model. From [22].

# 5 THz time-domain magneto-spectroscopy of a high-mobility two-dimensional electron gas

Strong free-carrier Faraday effects due to CR have also been observed in 2DEGs [47-50]. They demonstrated giant Faraday effects due to their high mobilities in very clean semiconductor heterostructures. However, not only simple enhancements of the classical Faraday effect, the high-mobility 2DEGs also allow one to observe qualitatively different phenomena, such as dephasing and quantum control.

## 5.1 Quantum Hall plateau in the THz frequency region

The Faraday effect is an optical analogue of the Hall effect. A natural question arises as to whether the quantum Hall effect survives in the AC regime. While DC properties of the integer quantum Hall effect is well understood, dynamic properties are not. Ikebe *et al*. examined this point experimentally by using THz magneto-spectroscopy [23]. Figure 9a shows the Faraday rotation angle and ellipticity of a high-mobility 2DEG in a GaAs/AlGaAs single heterojunction at 3 K. The data look similar to the Faraday parameters shown in Fig. 5 for *n*-InSb but show a very small width of the resonance due to the high mobility. The results basically obey the Drude model (solid lines in Fig. 9a), but the authors revealed slight deviation in the hatched region in Fig. 9a. To clarify the deviation, the authors deduced the THz Hall conductivity (off-diagonal components of the conductivity) from the Faraday rotation spectrum. They further normalized the conductivity to compensate for the frequency dependence due to CR and obtained normalized THz Hall conductivity as a function of Landau-level filling factor $\nu$ (blue solid circles in Fig. 9b). In Fig 9b, the normalized DC Hall conductivity (red solid line) of the sample is also shown as well as the theoretical ones for the classical Drude limit (dotted line) and the quantum Hall limit (dashed line). From this plot, not only the normalized DC Hall conductivity, but also the normalized THz conductivity deviates from the classical limit and exhibits a plateau-like behavior around $\nu = 2$ (expanded in Fig. 9c). At a higher temperature of 20 K, where the quantum Hall behavior is not expected, the plateau-like structure disappears and follows the classical Drude limit (green open squares). The results confirm a theoretical prediction by Morimoto *et al.* [51] and shed new light on the carrier dynamics in the quantum Hall regime.

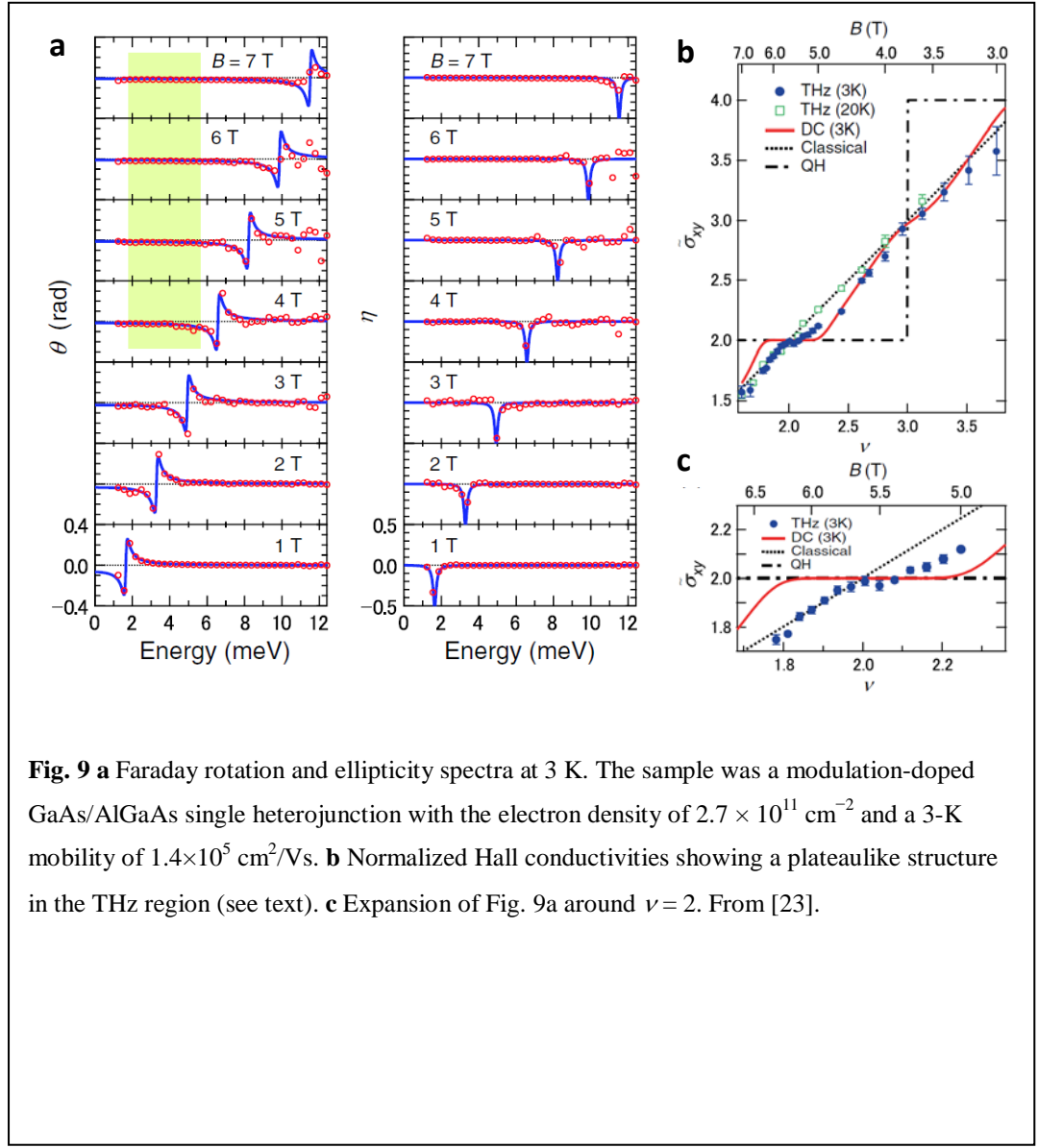

**Fig. 9 a** Faraday rotation and ellipticity spectra at 3 K. The sample was a modulation-doped GaAs/AlGaAs single heterojunction with the electron density of $2.7 \times 10^{11}$ cm$^{-2}$ and a 3-K mobility of $1.4 \times 10^5$ cm$^2$/Vs. **b** Normalized Hall conductivities showing a plateaulike structure in the THz region (see text). **c** Expansion of Fig. 9a around $\nu = 2$. From [23].

## 5.2 Coherent control of inter-Landau-level transitions

In contrast to the bulk systems, due to quantum confinement, the energy density of states of 2DEG becomes discrete (Landau levels separated by $\hbar\omega_c$), as in atomic systems, under the application of a strong perpendicular magnetic field. At high magnetic fields and low temperatures, where $k_BT < \hbar\omega_c$ ($k_B$ is the Bolzmann constant and $\hbar$ is the reduced Planck constant), CR is more appropriately viewed as a quantum transition between adjacent Landau levels. Here we describe classical (CR) and quantum mechanical (inter-Landau-level transition) pictures of the data obtained by THz time-domain magneto-spectroscopy of a high-mobility 2DEG and show coherent control of the inter-Landau-level transition using coherent THz pulses [24].

The left graph in Fig. 10a shows the *x*- and *y*-component of the transmitted THz pulses with [$E(B)$] and without [$E(0)$] a magnetic field of 2 T obtained with a similar THz magneto-spectroscopy system described in Section 4.1.  As clearly shown in the bottom traces [$E(B) - E(0)$], decaying sinusoidal oscillations with CR frequency $\omega_c/2\pi = eB/2\pi m^*$ of 0.816 THz are induced by the external magnetic field.  Here, $m^* = 0.068 m_0$ is the effective mass of electrons in GaAs.  The amplitude ratio of $E(B)$ and $E(0)$ in the frequency domain (Fig. 10a inset) has a dip at $\omega_c/2\pi$, showing that the magnetic field induced oscillation is due to CR absorption in the classical picture.  Figure 10b shows a parametric plot of induced oscillations from 7.85 to 9.05 ps (almost one full period of cyclotron oscillation, 1/0.816 THz = 1.22 ps) and it clearly shows that the absorbed THz wave is the circularly polarized CRA mode (circular anisotropy).  The reversed direction of the rotation in negative *B* (-2 T) further confirms this point.

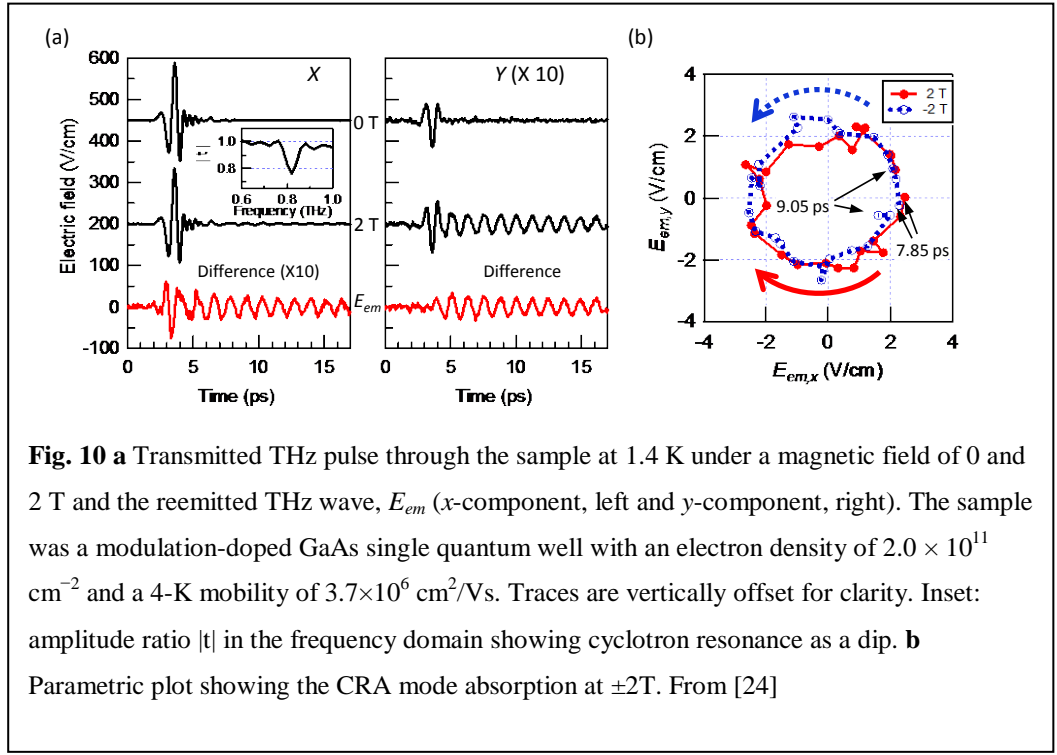

**Fig. 10 a** Transmitted THz pulse through the sample at 1.4 K under a magnetic field of 0 and 2 T and the reemitted THz wave, $E_{em}$ (*x*-component, left and *y*-component, right). The sample was a modulation-doped GaAs single quantum well with an electron density of $2.0 \times 10^{11}$ cm$^{-2}$ and a 4-K mobility of $3.7 \times 10^6$ cm$^2$/Vs. Traces are vertically offset for clarity. Inset: amplitude ratio |t| in the frequency domain showing cyclotron resonance as a dip. **b** Parametric plot showing the CRA mode absorption at ±2T. From [24]

A quantum mechanical picture is given as follows. At 2 T, the Landau level filling factor is 4.1, i.e., the lowest-two Landau levels (|0> and |1>) are completely filled (including the spin degeneracy), the next level |2> is almost empty, and all higher levels are empty. Due to the selection rules [52, 38] and the weak THz field employed, only a small population is excited from |1> to |2>, which enables us to treat the system as a two-level system. The incident THz pulse creates a coherent superposition of the two Landau levels with an oscillating polarization, which re-emits a coherent THz wave. In this picture, CR is the destructive interference of the incident THz pulse $E(0)$ and the re-emitted THz wave $E_{em}$, i.e., $E(B) = E(0) + E_{em}$. Thus, the *B*-induced oscillations (bottom traces in Fig. 10a) are exactly the re-emitted THz wave, $E_{em} = E(B) - E(0)$.

Next, Fig. 11 shows a coherent control of the quantum mechanical state of the two-level system with double pulse irradiation. At the arrival time of the second THz pulse ($t_{2nd} = 12.5$ ps, vertical broken line in Fig.11), the re-emitted THz waves under 2.05 and 2.175 T have phases of odd- and even-integer multiple of $\pi$, respectively. This phase difference of $\pi$ results in the opposite effect of the second THz pulse, i.e., quenching (2.05 T) and enhancing (2.175 T) of the re-emission. These two different situations are easily understood in terms of dynamics on the Bloch sphere (Fig. 11, right). In the case of quenching, the phase of odd-integer multiple of $\pi$ at $t_{2nd}$ means that the Bloch vector is at $(\theta_1,\pi)$ and the rotation operation about the axis $r_1$ by the second THz pulse moves the vector back to |1>. In the enhancement case, at $t_{2nd}$ the Bloch vector is at $(\theta_1, 0)$, which is the starting point of precession, and the rotation operation further increases $\theta$. These results show that an arbitrary quantum control of the Landau level superposition state is possible using coherent THz pulses. This phase-sensitive behaviour is well described with the optical Bloch equations (black curve in Fig. 11) by treating the system as a collection of independent single-electron two-level systems. This implies the absence of an influence of electron-electron interactions on CR frequencies, which is known as Kohn's theorem [53]. Coherent time-domain measurements broaden and extend the realm of applicability of Kohn's theorem to the coherent regime.

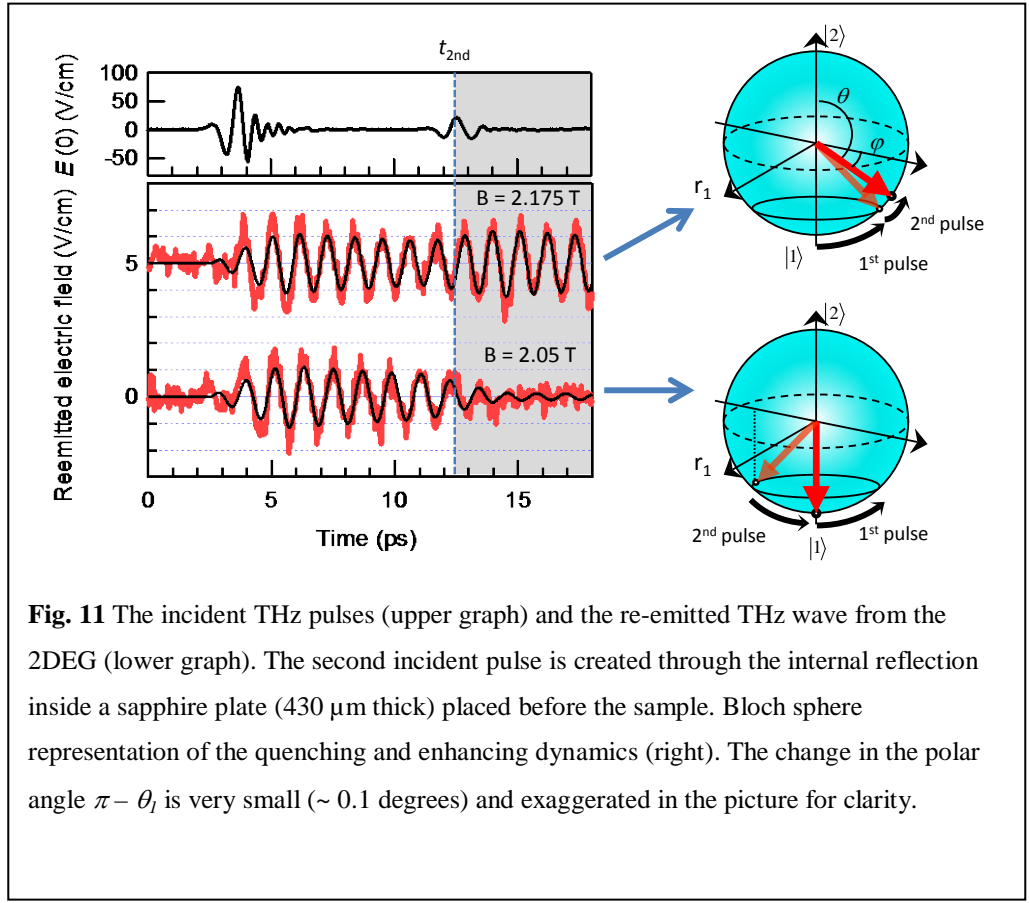

**Fig. 11** The incident THz pulses (upper graph) and the re-emitted THz wave from the 2DEG (lower graph). The second incident pulse is created through the internal reflection inside a sapphire plate (430 µm thick) placed before the sample. Bloch sphere representation of the quenching and enhancing dynamics (right). The change in the polar angle $\pi - \theta_1$ is very small (~ 0.1 degrees) and exaggerated in the picture for clarity.

## 6 Summary

In this article, we reviewed recent experiments on several anisotropic material systems performed in the THz frequency regime. The extreme linear dichroism in macroscopically-aligned carbon nanotubes enables one to make excellent THz linear polarizers. The giant circular dichroism and birefringence in *n*-InSb,

HgTe, and graphene, observed as the Faraday effect, have promising characteristics for constructing THz circular polarizers, wave plates, isolators, and modulators. Fundamental physics of carrier dynamics in 2DEGs in the quantum Hall regime are also addressed through the Faraday effect measurements. Further fundamental studies and device applications utilizing material anisotropies are anticipated in the THz frequency range.